\documentclass[twocolumn, switch]{article} 

\usepackage{preprint}
\usepackage{stfloats}
\usepackage{placeins}

\usepackage{algorithm}
\usepackage{algorithmic}
\usepackage{amsmath, amsthm, amssymb, amsfonts}

\usepackage[numbers,square]{natbib}
\usepackage[utf8]{inputenc}	
\usepackage[T1]{fontenc}	
\usepackage{xcolor}		
\usepackage[colorlinks = true,
            linkcolor = purple,
            urlcolor  = blue,
            citecolor = cyan,
            anchorcolor = black]{hyperref}	
\usepackage{booktabs} 		
\usepackage{nicefrac}		
\usepackage{microtype}		
\usepackage{lineno}		
\usepackage{float}			

\usepackage{lipsum}		

\usepackage{newfloat}
\DeclareFloatingEnvironment[name={Supplementary Figure}]{suppfigure}
\usepackage{sidecap}
\sidecaptionvpos{figure}{c}

\usepackage{titlesec}
\titlespacing\section{0pt}{12pt plus 3pt minus 3pt}{1pt plus 1pt minus 1pt}
\titlespacing\subsection{0pt}{10pt plus 3pt minus 3pt}{1pt plus 1pt minus 1pt}

\titleformat{\section}
  {\bfseries\large}            
  {\thesection}                
  {0.5em}
  {}
\titleformat{\subsection}
  {\bfseries\normalsize}
  {\thesubsection}
  {0.5em}
  {}

\title{\textbf{EAST: Environment-Aware Stylized Transition Along the Reality-Virtuality Continuum}}

\usepackage{titling}
\usepackage{orcidlink}
\usepackage{footmisc}
\author{
\textbf{Xiaohan Zhang},
\textbf{Kan Liu},
\textbf{Yangle Liu},
\textbf{Fengze Li},
\textbf{Jieming Ma},
\textbf{Yue Li}\thanks{Corresponding author}
}

\date{
Xi’an Jiaotong-Liverpool University, Suzhou, China\\[4pt]
\small
\{Xiaohan.Zhang2203, Kan.Liu23, Yangle.Liu22\}@student.xjtlu.edu.cn\\
fengzeli@liverpool.ac.uk,\ Jieming.Ma@xjtlu.edu.cn,\ Yue.Li@xjtlu.edu.cn
}

\begin{document}

\maketitle
\thispagestyle{empty}

\begin{abstract}
In the Virtual Reality (VR) gaming industry, maintaining immersion during real-world interruptions remains a challenge, particularly during transitions along the reality-virtuality continuum (RVC). Existing methods tend to rely on digital replicas or simple visual transitions, neglecting to address the aesthetic discontinuities between real and virtual environments, especially in highly stylized VR games. This paper introduces the Environment-Aware Stylized Transition (EAST) framework, which employs a novel style-transferred 3D Gaussian Splatting (3DGS) technique to transfer real-world interruptions into the virtual environment with seamless aesthetic consistency. Rather than merely transforming the real world into game-like visuals, EAST minimizes the disruptive impact of interruptions by integrating real-world elements within the framework. Qualitative user studies demonstrate significant enhancements in cognitive comfort and emotional continuity during transitions, while quantitative experiments highlight EAST's ability to maintain visual coherence across diverse VR styles.
\end{abstract}
\vspace{0.35cm}


\section{Introduction}
In recent years, immersive virtual reality (VR) gaming has transformed the entertainment industry by delivering deeply engaging and interactive experiences. A key differentiator of VR gaming is its ability to fully immerse users in virtual environments, setting it apart from other gaming platforms. However, real-world interruptions—such as phone calls, notifications, or physical interactions—pose significant challenges to maintaining this immersion \cite{freina2015literature, slater2016enhancing}. These disruptions force abrupt shifts of attention from the virtual to the real world, fragmenting the experience and undermining the seamless engagement VR aims to provide. Immersion in VR is a multidimensional concept encompassing sensory, cognitive, and emotional dimensions \cite{szabo2019interaction}. Sensory immersion relates to how convincingly the VR system stimulates the user’s senses to recreate a virtual world. Cognitive immersion involves the user's mental engagement and sustained focus, while emotional immersion reflects the depth of their emotional connection to the VR content \cite{monteiro2021hands}. Real-world interruptions disrupt all these dimensions, breaking the continuity of the virtual experience and forcing users to reorient themselves to their physical surroundings \cite{evans2020hermeneutic}.

This challenge aligns with the broader framework of the Reality-Virtuality Continuum (RVC) \cite{milgram1994taxonomy}, which spans from fully real environments to fully virtual ones, with intermediate stages such as augmented reality (AR) and augmented virtuality (AV). While AR integrates virtual elements into the real world, AV introduces real-world elements into a predominantly virtual environment. Managing real-world interruptions within a VR gaming context involves navigating this continuum by blending real-world stimuli into the virtual environment without disrupting the user’s immersion.

Research has shown that such interruptions can significantly impact the user's sense of presence—the psychological state of being there in the virtual environment \cite{rahimi2018scene}. Studies have examined the effects of real-world distractions, such as a ringing phone, on users' cognitive and affective responses during virtual social interactions \cite{horst2024back}. These findings indicate that real-world distractions negatively affect recognition, recall, interpersonal liking, and communication satisfaction, highlighting the detrimental impact of interruptions on user engagement in VR environments. Moreover, the nature of the interruption plays a crucial role in the degree of disruption experienced. Active interruptions, where users must respond to real-world stimuli (e.g., answering a phone call), are more disruptive than passive ones, where users are merely aware of the distraction but do not need to act upon it \cite{pointecker2020exploration}. This distinction underscores the importance of developing strategies to manage real-world interruptions in VR, aiming to minimize their impact on user immersion.

The challenge lies in creating transition techniques that allow users to address real-world interruptions without completely disengaging from the virtual environment. Traditional methods, such as pausing the VR experience or fading to a neutral screen, often result in abrupt and jarring transitions that break immersion. There is a growing need for more sophisticated approaches that can seamlessly integrate real-world elements into the VR experience, maintaining aesthetic and perceptual coherence to preserve user engagement.

Traditional transition techniques in VR, such as fade, dissolve, translate, and combine, have been primarily designed to facilitate movement between virtual environments. While effective within the virtual domain, these methods often fail to address the significant aesthetic and perceptual disparities between stylized VR game environments and the real world. This limitation becomes particularly evident during real-world interruptions, where such transitions result in jarring experiences that disrupt user immersion. For instance, fade and dissolve transitions rely on gradually changing the opacity of virtual elements, creating smooth visual shifts within virtual environments. However, when applied to scenarios involving real-world interactions, these techniques produce abrupt shifts that lack contextual alignment. In this process, virtual elements from the replica environment are gradually introduced via dissolve effects, progressing toward contextual continuity by replicating real-world lighting. Despite these efforts, the stark contrast between the stylized VR game aesthetics and real-world settings remains a challenge. Intermediate states often fail to maintain perceptual coherence, as walls and floors fade while prominent objects dissolve, creating spatial inconsistencies and cognitive overload. The final stage materializes the target environment, but the transition still accentuates visual disparities that disrupt the continuity of the user experience.

A fundamental limitation of traditional techniques is their inability to harmonize the aesthetic differences between stylized VR environments and the real world. Stylized VR games often feature distinctive art styles, color palettes, and visual effects that deviate significantly from the appearance of the real world. For example, transitioning from a cel-shaded virtual environment to the real world using a simple fade or dissolve effect highlights the visual disparity, undermining the immersive experience. These perceptual mismatches create a sense of disconnection for users, particularly when abrupt transitions do not account for the unique artistic features of the VR environment. Research on immersive transitions underscores the importance of maintaining contextual and perceptual coherence during such shifts. Poorly managed transitions have been shown to cause disorientation, discomfort, and a diminished sense of presence, especially in scenarios involving head-mounted displays (HMDs). For instance, studies highlight how mismatches in visual and spatial cues during transitions lead to reduced cognitive comfort and engagement, underscoring the need for more nuanced approaches to cross-reality transitions.

Current transition methods inadequately address these challenges, focusing predominantly on screen-driven visual effects without considering the stylistic linkage between the displayed virtual content and the real-world environment. This oversight is especially pronounced in VR gaming, where the aesthetic gap between highly stylized game environments and the real world is substantial. Existing techniques lack mechanisms to integrate real-world elements in a manner that preserves the artistic coherence of the virtual scene. Consequently, users experience a perceptual and cognitive disconnect that disrupts immersion, particularly during moments of real-world interaction. Addressing this aesthetic and perceptual gap requires novel approaches that not only ensure smoother transitions but also maintain the stylistic integrity of the VR environment. As highlighted in recent literature, achieving such transitions demands a framework that effectively blends game aesthetics with real-world contexts, thereby reducing the perceptual disparity and enhancing user comfort.

Existing research has extensively examined various aspects of transitioning between VR environments \cite{feld2023keep, husung2019portals}, focusing on enhancing user engagement through digital replicas \cite{steinicke2009does} and managing transitions across different stages of the RVC \cite{billinghurst2001magicbook, sra2016procedurally, feld2024simple}. While these studies have contributed significantly to understanding transitions within VR, they predominantly rely on static replicas or simplistic visual effects, which are insufficient for addressing the aesthetic and perceptual challenges posed by highly stylized VR gaming environments. Most approaches aim to provide linear transitions between stages of the RVC, such as augmented reality (AR), augmented virtuality (AV), and fully virtual environments, but they fail to consider the stylistic disparities and cognitive challenges associated with abrupt context switching during real-world interruptions. A comprehensive framework for transitions must integrate replicas with advanced transition techniques that harmonize the aesthetic and perceptual coherence across all stages of the RVC. Such a framework would facilitate seamless navigation between real-world and virtual environments, minimizing the cognitive disruption caused by real-world interruptions. Gradually incorporating virtual elements into the real-world context can help achieve more natural transitions, progressively blending AR and AV into the process. By aligning real-world elements with the artistic features of VR games, this approach can significantly reduce the perceptual gap between the real and virtual worlds. Despite the progress in RVC-related research, existing methods fall short in maintaining aesthetic coherence, particularly in the context of stylized VR games, where visual disparities are most pronounced. For users with limited experience using head-mounted displays (HMDs), poorly managed transitions exacerbate discomfort and diminish the immersive quality of the experience. Addressing these limitations requires a shift toward dynamic and stylistically adaptive frameworks that not only preserve the artistic integrity of the VR environment but also enhance cognitive comfort and user engagement during transitions.

\begin{figure*}[!t]
    \centering
    \includegraphics[width=1\textwidth]{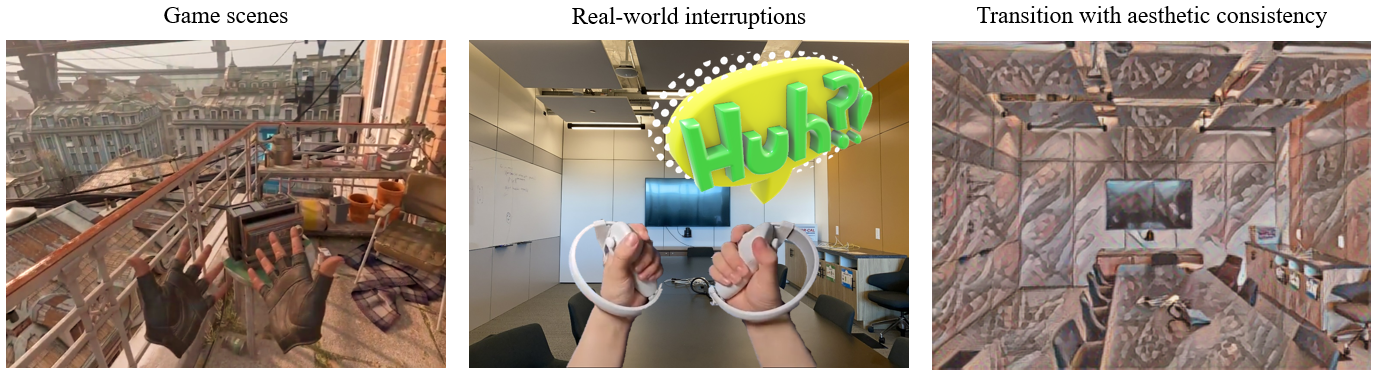}
        \caption{The implem
ented environment-aware stylized transition with seamless aesthetic consistency. From left to right: participants' first-person perspective of the the game scene; a real-world scenario with an interruption; the transition from VR to the real world with aesthetic consistency.}   
    \label{fig:1}
\end{figure*}

To address the aforementioned limitations, we propose a novel framework based on stylized 3DGS to seamlessly integrate real-world interruptions into stylized VR environments (see Fig.~\ref{fig:1}). Our approach leverages advanced 3D reconstruction techniques and neural style transfer to visually align real-world objects with the aesthetics of the virtual environment. For instance, a ringing phone in the real world can be transformed to match the artistic style of the VR game, ensuring that users can address real-world tasks without breaking their immersion. This integration significantly reduces the cognitive load associated with abrupt context switching, creating a seamless and coherent user experience. By maintaining aesthetic consistency, our framework enhances the user's sense of presence, addressing the perceptual disparities that disrupt immersion in existing transition methods.

Recent advancements in 3DGS have demonstrated its exceptional potential for real-time, high-fidelity 3D scene 
reconstruction. By modeling objects as collections of 3D Gaussians, 3DGS achieves geometric precision while maintaining computational efficiency, making it particularly well-suited for the demanding requirements of VR applications. The technique's ability to represent complex surfaces with continuous and smooth representations enables detailed reconstructions of real-world objects with minimal artifacts. When coupled with neural style transfer, 3DGS extends its capabilities to incorporate the stylistic features of VR environments. Neural style transfer uses deep neural networks to extract stylistic patterns, such as textures, colors, and shapes, from a reference style and apply them to reconstructed 3D objects. This combination ensures that real-world elements, when introduced into the VR scene, align seamlessly with the virtual world's visual language. By building upon these advancements, our framework bridges the gap between the real and virtual worlds, offering a robust solution to the aesthetic mismatches that hinder current RVC frameworks.

Our proposed framework fundamentally extends the capabilities of the RVC by addressing aesthetic disparities and enabling fluid transitions across its stages, from fully real to fully virtual environments. Existing RVC models focus primarily on static replicas or linear transitions, neglecting the dynamic and stylistic integration required for immersive VR experiences. By incorporating stylized 3DGS, we enhance the aesthetic coherence and perceptual alignment of transitions, ensuring that real-world interruptions, such as objects or notifications, blend naturally into the virtual environment without disrupting immersion. This approach transforms previously disruptive interactions into integrated components of the VR experience. The novel contributions of our work include: 
\begin{enumerate}
    \item This paper introduces the Environment-Aware Stylized Transition (EAST) framework, leveraging a novel style-transferred 3D Gaussian Splatting (3DGS) technique to seamlessly integrate real-world interruptions into virtual environments, preserving aesthetic consistency.
    \item EAST addresses the aesthetic disparities between stylized VR game environments and the real world, optimizing transitions along the reality-virtuality continuum (RVC) with a focus on maintaining visual coherence despite significant style differences.
    \item Quantitative experiments validate the framework’s efficacy, demonstrating superior structural similarity and real-time rendering performance, ensuring smooth and efficient transitions during game-play.
    \item Qualitative user studies confirm that EAST significantly enhances cognitive comfort, emotional continuity, and overall immersion, improving user satisfaction during real-world interruptions in VR gaming scenarios.
\end{enumerate}

\section{Method}
\begin{figure*}[!t]
    \centering
    \includegraphics[width=1\textwidth]{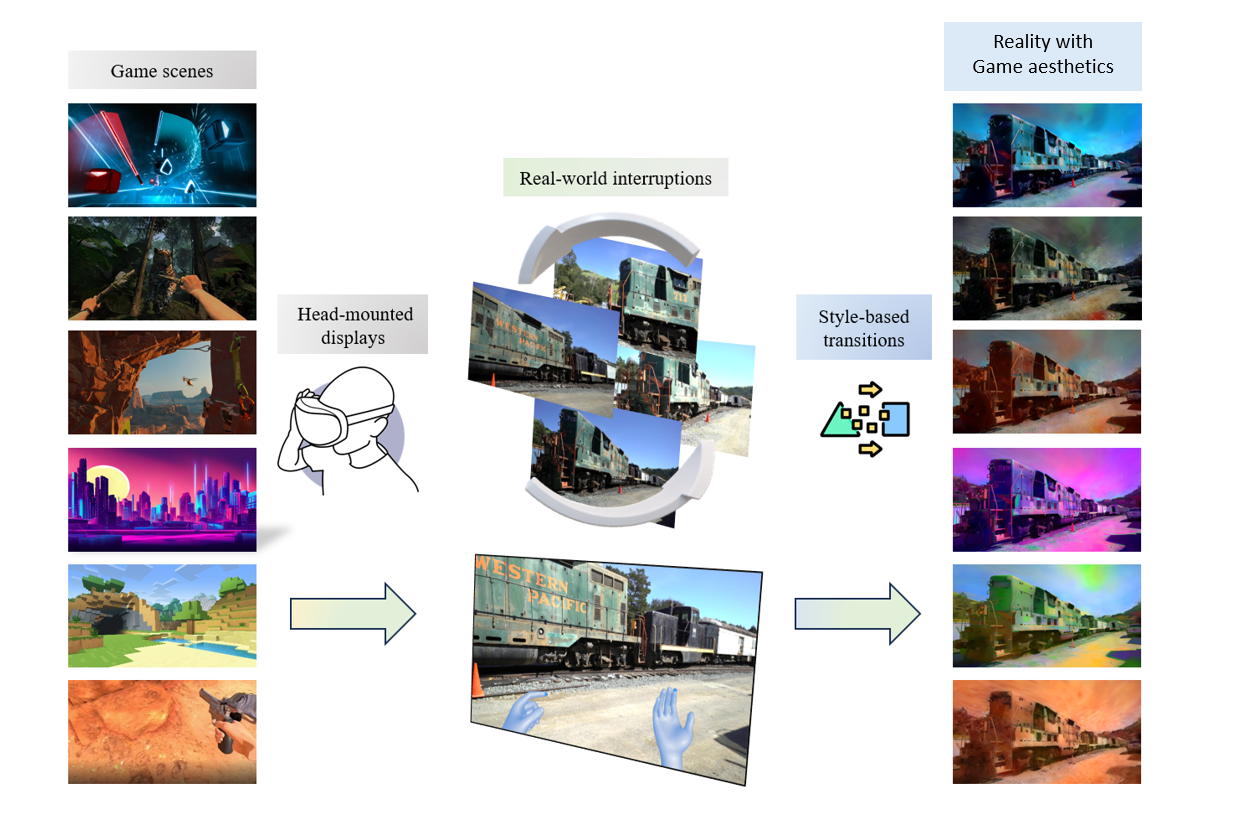}

        \caption{Overview of our environment-aware stylized transition (EAST) approach. Real-world interruptions are seamlessly integrated into VR by stylizing objects to match the game’s aesthetics, maintaining immersion and minimizing disruption.}   
    \label{233.png}
\end{figure*}

Our approach utilizes 3DGS \cite{kerbl20233d} and RVC frameworks to achieve an efficient and visually consistent style transfer between a 2D style image and a 3D scene. Fig.~\ref{233.png} provides an overview of our style-based transition method. The input to the system includes multiview images and corresponding structure-from-motion points, which serve as the basis for reconstructing the 3D scene. The multiview images are processed through a 3DGS pipeline, enabling the generation of a detailed and geometrically accurate scene representation. Meanwhile, the style image undergoes feature extraction via VGG \cite{simonyan2014very}, which captures its visual characteristics.

\begin{figure*}[!t]
    \centering
    \includegraphics[
width=1\textwidth]{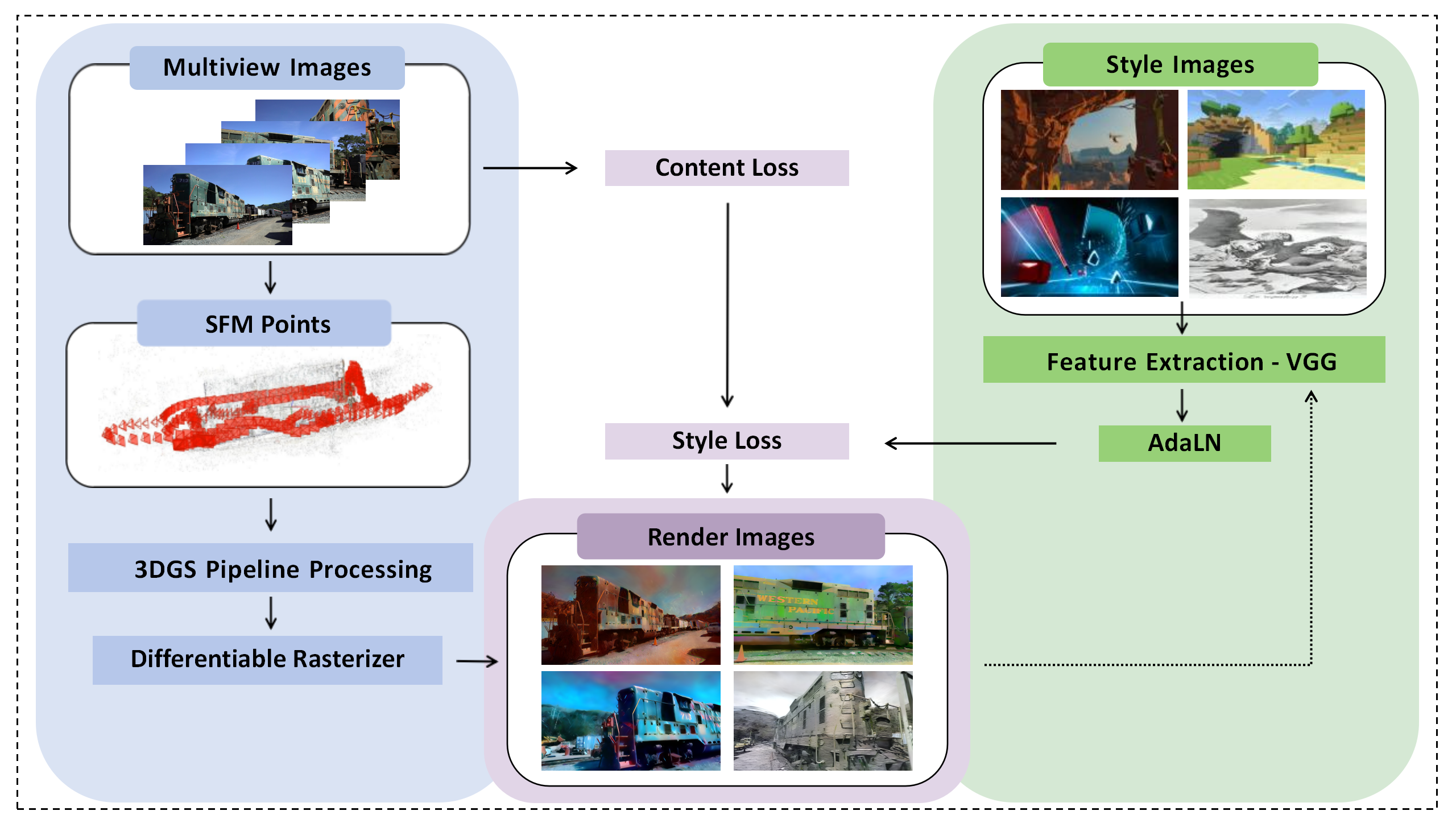}
        \caption{The pipeline of our method for achieving style transfer in the reconstructed 3D environment.}   
    \label{liuchengtu.png}
\end{figure*}

In the style transfer process, the extracted features from the style image are ali
gned with the 3D scene representation. The adaptive layer normalization (adaLN) \cite{perez2018film} is employed to modulate the style characteristics onto the 3D scene, guided by the perceptual losses: content loss and style loss. These losses ensure that the generated scene retains its original content while being transformed according to the target style. Finally, the stylized scene is rendered through a differentiable rasterizer, resulting in a high-quality image that reflects both the geometric accuracy of the 3D reconstruction and the aesthetic details from the style image. The pipeline of our method is illustrated in Fig.~\ref{liuchengtu.png}.

\subsection{Preliminaries: 3DGS} \label{sec:3dgs}

 Gaussian splatting encapsulates 3D scene information using a set of 3D colored Gaussians. This technique exhibits rapid inference speeds and exceptional reconstruction quality compared to NeRF. 
To represent the scene, each Gaussian is described by a centroid $p = \{x, y, z\} \in \mathbb{R}^3$, a 3D vector $s \in \mathbb{R}^3$ for scaling, and a quaternion $q \in \mathbb{R}^4$ for rotation. Additionally, an opacity value $\alpha \in \mathbb{R}$ and a color vector $c$ represented in the coefficients of a spherical harmonic (SH) function of degree 3 are used for fast alpha-blending during rendering.
These trainable parameters are collectively symbolized by
$G_{\theta_i}$, where $G_{\theta_i} = \{p_i, s_i, q_i, \alpha_i, c_i\}$, representing the parameters for the $i$-th Gaussian. To visualize the 3D Gaussians and supervise their optimization, 3DGS projects them onto the 2D image plane. The implementation leverages differentiable rendering and gradient-based optimization on each pixel for the involved Gaussians. The pixel color $c^{\alpha}$ is determined by blending the colors $c_i$ of those ordered Gaussians that overlap the pixel. This process can be formulated as:
\begin{equation}
    c^{\alpha}=\sum_{i\in N}T_i \alpha_i c_i
\end{equation}
where $T_i$ is the accumulated transmittance, and $\alpha_i$ is the alpha-compositing weight for each Gaussian. This method ensures high-quality scene reconstruction with fast inference speeds, forming the basis for stylized 3D reconstruction in the EAST framework.

\subsection{Style Transfer} \label{sec:style_transfer}
To achieve style transfer in the context of our 3D scene reconstruction, we employ a two-step approach: feature extraction and adaptive style modulation. The process begins by extracting features from the 2D style image using the VGG network \cite{simonyan2014very}, which captures the perceptual characteristics of the style image. Specifically, the VGG network's convolutional layers are used to extract deep features that represent the color, texture, and structure of the target style.

Once the features are extracted, we apply AdaIN \cite{perez2018film} to modulate the style characteristics onto the 3D scene. AdaIN works by normalizing the feature maps of the 3D scene based on the style features from the style image, and then adapting these normalized values with the style statistics (mean and variance) derived from the VGG network’s output. The transformation can be mathematically expressed as:
\begin{equation}
\hat{y}_i = \gamma_i \left( \frac{x_i - \mu_x}{\sigma_x} \right) + \beta_i
\end{equation}
where \( x_i \) is the feature map of the 3D scene at layer \( i \), \( \mu_x \) and \( \sigma_x \) are the mean and standard deviation of the scene’s features, and \( \gamma_i \) and \( \beta_i \) are the learned style parameters derived from the style image's feature map. This modulation enables the scene to adopt the visual appearance of the style image while retaining the geometric content of the original 3D scene.

The style transfer process is further guided by two perceptual losses: content loss and style loss. Content loss ensures that the generated 3D scene retains its original content by minimizing the difference between the content features of the stylized 3D scene and the ground truth. This is typically calculated using L2 loss:
\begin{equation}
L_{\text{content}} = \| F_{\text{content}} - F_{\text{gt}} \|^2
\end{equation}
where \( F_{\text{content}} \) and \( F_{\text{gt}} \) are the feature maps of the generated scene and the ground truth image, respectively. Style loss, on the other hand, ensures that the stylized scene reflects the visual characteristics of the style image. It is computed using the Gram matrix \cite{gatys2015neural} of the style features, which captures the correlations between different feature maps. The style loss is given by:
\begin{equation}
L_{\text{style}} = \| G(F_{\text{generated}}) - G(F_{\text{style}}) \|^2
\end{equation}
where \( G(F) \) is the Gram matrix of feature map \( F \), representing the correlations between different features. The total loss for style transfer is then a weighted sum of the content and style losses:
\begin{equation}
L_{\text{total}} = w_c L_{\text{content}} + w_s L_{\text{style}}
\end{equation}
where \( w_c \) and \( w_s \) are the weights for content and style losses, respectively. The optimization process minimizes this total loss to generate a stylized 3D scene that preserves both content and style features.

\subsection{Differentiable Rasterizer} \label{sec:differentiable_rasterizer}
The differentiable rasterizer plays a critical role in enabling gradient-based optimization for 3D scene reconstruction. After the style features are applied to the 3D scene using AdaIN, the scene is rendered into a 2D image for comparison against the ground truth. The rendering process is differentiable, meaning that the gradients of the loss functions with respect to the scene’s parameters, including the Gaussian splats, rotation, scaling, and opacity, can be computed and used to update the scene representation during training.

In this framework, the rendered image is obtained by summing the contributions of overlapping Gaussians in the scene, where each Gaussian is represented by its color \( c_i \), opacity \( \alpha_i \), and other properties. The final rendered pixel color \( c^{\alpha} \) is computed as a weighted sum of these Gaussians, as shown in the following formula:
\begin{equation}
c^{\alpha} = \sum_{i \in N} T_i \alpha_i c_i
\end{equation}
where \( T_i \) is the accumulated transmittance, and \( \alpha_i \) is the alpha-compositing weight of the \( i \)-th Gaussian. The differentiable nature of this rasterization allows the optimization process to propagate gradients through the rendering step, facilitating the adjustment of the 3D scene parameters during training.

The use of a differentiable rasterizer enables a seamless integration of the 3D scene and style transfer, allowing the entire pipeline to be trained end-to-end. The gradients from the content and style losses are back-propagated through the rasterizer to update the scene’s Gaussian parameters, ensuring that both geometric accuracy and style characteristics are optimized.

\subsection{User Interface}
To facilitate interaction with the system and enable real-time monitoring of the style transfer process, we incorporate a user interface (UI) based on socket communication. The UI provides a way to visualize and control the various stages of the style transfer and 3D reconstruction process, including adjusting camera parameters, scaling, and viewing the intermediate results during training.

The UI script acts as a bridge between the user and the system, enabling communication between the training process and the user’s interface. It manages the camera parameters and scene transformations, allowing users to manipulate the viewpoint and interact with the 3D scene in real-time. The UI also provides feedback on the current status of the training, including loss values and the progress of the style transfer.

The system’s architecture involves a client-server model, where the server performs the heavy computational tasks while the client communicates with the server to request updates and send control signals. The socket communication listens for commands from the user and responds with updated scene renders. This setup enables the user to interact with the 3D scene, view intermediate renders, and monitor the training process without interrupting the workflow. The user interface enhances the interactivity of the system, providing real-time control and monitoring of the style transfer process, thus improving the user experience and facilitating easier experimentation with different style transfer configurations.

\section{Experiments} 
We include the implementation details, especially the training settings, in our supplementary document. The following experiments focus on evaluating the accuracy of stylization under the context of style-based transitions.

\subsection{Datasets}
We conduct extensive experiments to evaluate our style-based transitions approach using multiple real-world scene datasets. Specifically, we use the LLFF dataset~\cite{mildenhall2019local}, which contains forward-facing scenes, and the Tanks \& Temples (T\&T) dataset~\cite{knapitsch2017tanks}, which includes large-scale unbounded 360$^\circ$ outdoor scenes. 

\begin{figure*}[!t]

    \includegraphics[width=0.95\textwidth]{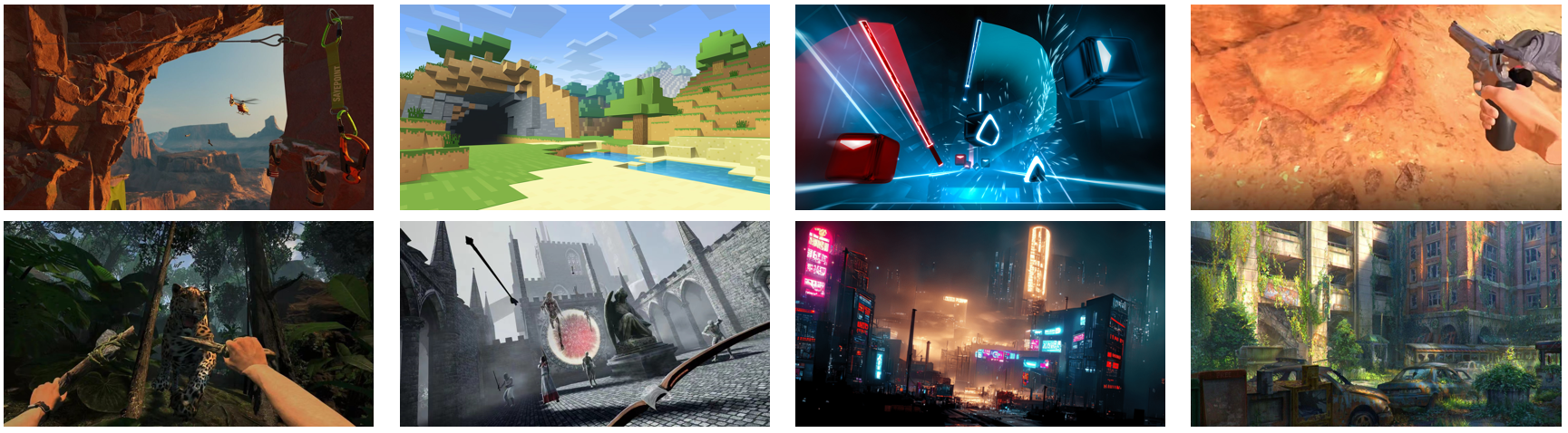}
        \caption{Sample images from the game-style dataset. These screenshots, collected from various VR and streamable games, represent the diverse styles used in our experiments to evaluate transitions between game aesthetics and real-world environments.}
    \label{fig:d1}
\end{figure*}

For game-related stylization, we collected a variety of VR games and other streamable content from gaming devices to form our game-style dataset as shown in Fig. \ref{fig:d1}. This game-centric dataset allows us to better address the need for transitioning from game aesthetics to real-world environments. We collected a total of 20 different games, and for each game, we randomly captured 10 screenshots at different stages, creating a comprehensive and diverse set of gaming-style examples. These datasets enable us to rigorously test the effectiveness and robustness of our style-based transitions framework across various environments, styles, and gaming scenarios.

\subsection{Baselines}
To 
thoroughly evaluate the performance of our method in the context of style-based transitions, we compare it with three state-of-the-art 3D stylization methods: LSNV~\cite{huang2021learning}, StyleRF~\cite{liu2023stylerf}, and StyleGaussian~\cite{liu2024stylegaussian}. These methods utilize different 3D representations, where LSNV is based on point clouds, StyleRF is based on NeRF, and StyleGaussian is based on 3DGS. While all three methods are feed-forward-based, they differ in their handling of 3D scenes and stylization processes.

For a fair comparison, we utilize the official code and pre-trained models released by the authors of these methods. We evaluate all methods using three key metrics: SSIM to assess structural similarity, LPIPS for perceptual distance, and DISTS for evaluating detail preservation. These metrics allow us to benchmark the performance of our style-based transitions approach against existing techniques, focusing on style accuracy, consistency, and smoothness of transitions between game aesthetics and real-world environments.

To evaluate the effectiveness of our style-based transitions framework, we performed extensive quantitative comparisons using our proposed game-style dataset. The dataset provides a rich variety of stylized exemplars to rigorously test the adaptability and robustness of our approach. Our results are compared with three state-of-the-art 3D stylization methods: LSNV~\cite{huang2021learning}, StyleRF~\cite{liu2023stylerf}, and StyleGaussian~\cite{liu2024stylegaussian}, each representing a different 3D representation strategy. As shown in Table~\ref{tab:comp}, we measure the performance of these methods using three key metrics: SSIM to evaluate structural similarity, LPIPS to assess perceptual similarity, and DISTS to measure the preservation of details during stylization transitions.

\subsection{Quantitative Comparisons} 
\begin{table}[t]
\centering
\caption{Quantitative comparisons on stylization under novel views using our proposed game-style dataset. We report SSIM ($\uparrow$), LPIPS ($\downarrow$), and DISTS ($\downarrow$) for our method and selected baselines.}
\begin{tabular}{cccc}
    \toprule
    Metrics & SSIM ($\uparrow$) & LPIPS ($\downarrow$) & DISTS ($\downarrow$) \\
    \midrule
    LSNV & 0.13 & 0.65 & 0.33 \\
    StyleRF    & 0.41 & 0.45 & 0.30 \\
    StyleGaussian & 0.35 & 0.52 & 0.30 \\
    Ours       & \textbf{0.55} & \textbf{0.39} & \textbf{0.26} \\
    \bottomrule
\end{tabular}
\label{tab:comp}
\end{table}

Our method consistently outperforms the baselines across all metrics. Specifically, our approach achieves an SSIM of 0.55, significantly higher than StyleRF (0.41), StyleGaussian (0.35), and LSNV (0.13). This indicates that our method preserves the structural integrity of the original content more effectively, ensuring that the stylization does not distort the underlying scene geometry.

In terms of perceptual quality, as measured by LPIPS, our method achieves the lowest score of 0.39, compared to StyleRF (0.45), StyleGaussian (0.52), and LSNV (0.65). Lower LPIPS values indicate that our stylized outputs are more perceptually similar to the original scenes, preserving fine stylistic details while minimizing artifacts that could detract from the immersive experience, which is particularly critical in VR environments.

The DISTS metric, which measures detail preservation, further highlights the strength of our approach. Our method achieves a score of 0.26, outperforming StyleRF (0.30), StyleGaussian (0.30), and LSNV (0.33). This demonstrates that our style-based transitions effectively maintain visual details even during the transition between game aesthetics and real-world environments, which is crucial for providing a seamless and immersive user experience.

The visual results presented in the accompanying figure further illustrate these findings. As seen in the top half of the figure, our method excels in capturing intricate geometric details from the stylized game scenes while preserving semantic content and maintaining consistency across views. The baseline methods often fail to capture these intricate details, leading to either oversimplified or overly distorted transitions. For example, the StyleRF and StyleGaussian outputs exhibit visible blurring and loss of key stylistic features, especially in highly detailed scenes, such as the train environment and dinosaur skeleton transitions.

\begin{figure*}[!t]
    \centering
    \includegraphics[width=1\textwidth]{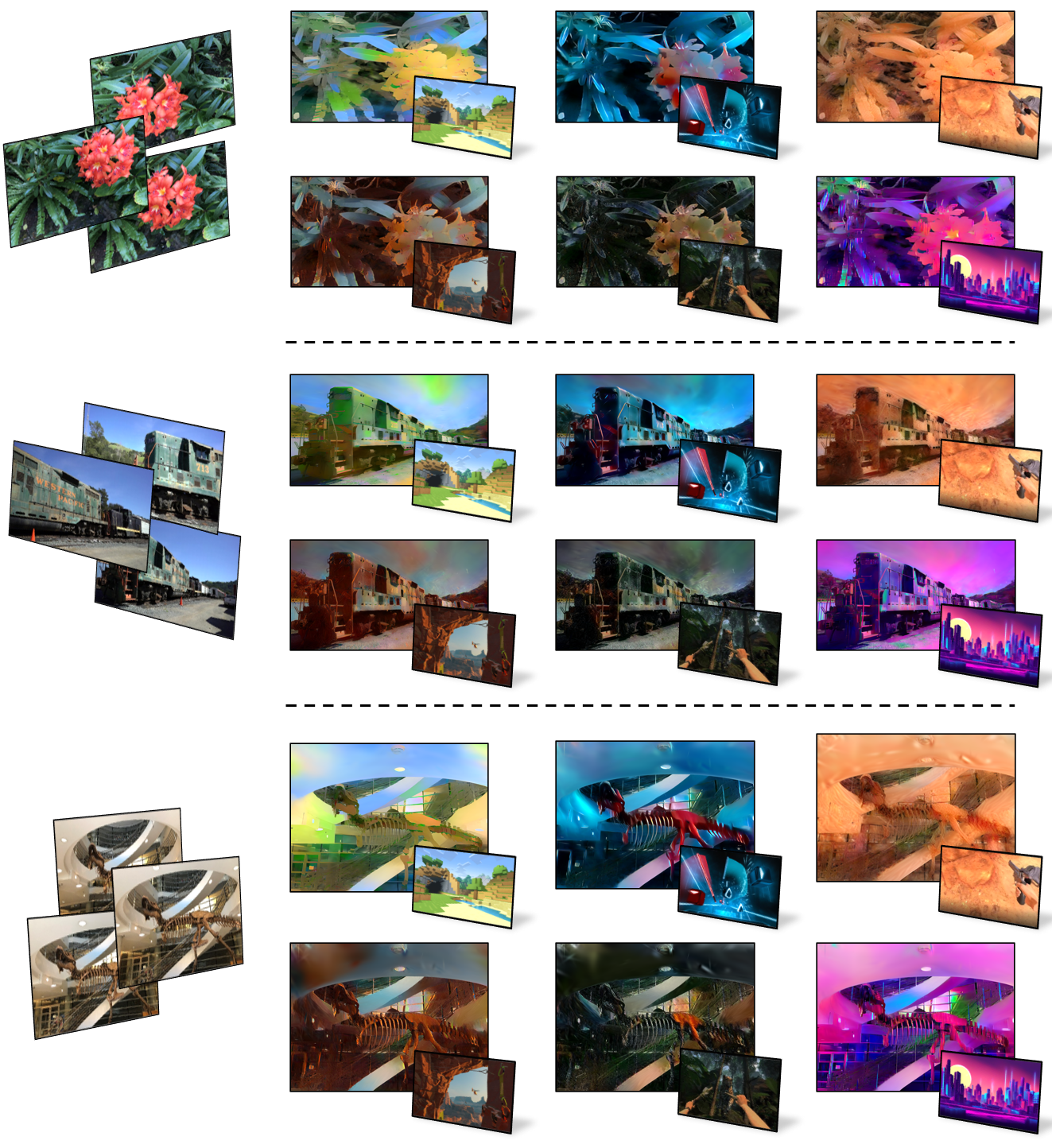}
        \caption{Demonstration of the style transfer results. Left: images of the real-world environment; Right: game-aware style transfer results based on the game scene. These screenshots, collected from various VR and streamable games, represent the diverse styles used in our experiments to evaluate transitions between game aesthetics and real-world environments.}
    \label{fig:d2}
\end{figure*}

In contrast, our approach, as shown in Fig.~\ref{fig:d2}, achieves superior style transfer with smoother, more visually coherent transitions, even for complex styles like those seen in VR gaming environments. By leveraging the richness of our game-style dataset, our method demonstrates its capability to handle diverse and highly detailed scenes, providing a more natural and immersive transition from virtual to real-world settings.

In summary, our method achieves improved results across key quantitative metrics but also demonstrates its effectiveness in producing visually coherent and detail-rich stylized transitions, as shown in the Fig.~\ref{fig:d2}. These results solidify the robustness of our style-based transitions framework in addressing the challenges of integrating game aesthetics with real-world environments.

\section{Performance in VR HMDs} \label{sec:performance}

Integrating our environment-aware stylized transition (EAST) approach within Virtual Reality Head-Mounted Displays (VR HMDs) demands meticulous consideration of both system performance and user experience. This section provides an in-depth analysis of the rendering performance, computational optimizations, latency management, and the resultant impact on user comfort and immersion. By addressing these facets, we demonstrate the practicality and effectiveness of our method in delivering seamless transitions within immersive VR environments.

\subsection{Rendering Performance and Latency}

Achieving real-time rendering is critical for maintaining immersion and preventing motion sickness in VR applications. Our EAST approach leverages Stylized 3D Gaussian Splatting (3DGS), optimized for high frame rates essential for VR HMDs.

\subsubsection{Frame Rate Analysis}

We conducted extensive benchmarking to evaluate the frame rates achieved by our EAST approach across various VR HMDs, including the PICO 4 Pro and Meta Quest 2. Table \ref{tab:framerate} summarizes the performance metrics observed during these tests.

\begin{table}[h]
\centering
\caption{Frame rate performance of EAST on various VR HMDs}
\begin{tabular}{lcc}
\toprule
\textbf{HMD Model} & \textbf{Without EAST} & \textbf{With EAST} \\
\midrule
PICO 4 Pro        & 90 FPS               & 85 FPS            \\
Meta Quest 2    & 72 FPS               & 68 FPS            \\
HTC Vive Pro      & 90 FPS               & 84 FPS            \\
\bottomrule
\end{tabular}
\label{tab:framerate}
\end{table}

The introduction of EAST incurs a modest frame rate reduction of approximately 5-7 FPS across different HMDs. Despite this slight decrease, the frame rates remain well within the optimal range for VR applications, ensuring a smooth and comfortable user experience. The minimal performance overhead is attributed to the efficient implementation of the 3DGS-based stylization pipeline, which prioritizes real-time processing without compromising visual fidelity. The maintained frame rate is consistent with previous studies, which suggest that a stable 70 FPS or higher is generally acceptable for ensuring a smooth VR experience \cite{10536453}.

\subsubsection{Latency Management}

Low latency is imperative to prevent disorientation and maintain the sense of presence in VR. Our approach employs several strategies to minimize latency:

\begin{itemize}
    \item \textbf{Asynchronous Processing}: Stylization tasks are offloaded to separate GPU threads, allowing the main rendering loop to operate without interruption. This decoupling ensures that real-time interactions within the VR environment remain responsive.
    \item \textbf{Optimized Data Transfer}: We utilize high-speed memory interfaces and optimized data transfer protocols (e.g., NVLink) to facilitate rapid movement of stylization data between the CPU and GPU. This reduces bottlenecks and ensures that stylized 3DGS models are available for rendering with minimal delay.
\end{itemize}

By implementing these latency reduction techniques, our EAST approach maintains high responsiveness, crucial for immersive VR experiences.

\subsection{Computational Optimizations}

To sustain real-time performance on VR HMDs, especially those with limited computational resources, our method incorporates several optimizations:

\subsubsection{Level of Detail (LOD) Management}

We implement Level of Detail (LOD) techniques to dynamically adjust the complexity of 3DGS models based on their distance from the user's viewpoint. Objects located farther from the user are rendered with fewer Gaussians, reducing computational load without significantly impacting perceived visual quality. This adaptive approach ensures that resources are allocated efficiently, prioritizing detailed rendering for objects within the user's immediate focus.

\subsubsection{Parallel Processing and GPU Acceleration}

Our stylization algorithm is designed to exploit the parallel processing capabilities of modern GPUs. By utilizing CUDA and OpenCL frameworks, we enable concurrent processing of multiple Gaussians, significantly accelerating the stylization pipeline. This parallelization allows for real-time application of complex style transfers without introducing noticeable delays.

\subsubsection{Efficient Memory Utilization}

Memory optimization is achieved through the compression of 3DGS representations and the use of efficient data structures. By reducing the memory footprint of 3D Gaussian models, we facilitate faster access and manipulation of Gaussian parameters, which is essential for maintaining high performance on devices with constrained memory resources. Additionally, we employ memory pooling and reuse strategies to minimize fragmentation and overhead.

\subsection{Integration with VR HMDs}

Our EAST approach is designed to be highly compatible with a wide range of VR HMDs, ensuring broad applicability and ease of deployment.

\subsubsection{Cross-Platform Compatibility}

Implemented using cross-platform graphics APIs such as Vulkan and DirectX, our method ensures compatibility with various VR systems, including PC-tethered and standalone HMDs. This flexibility allows developers to integrate EAST into diverse VR applications without significant modifications, facilitating widespread adoption.

\subsubsection{Scalable Stylization Quality}

Users can adjust the level of stylization based on their hardware capabilities and performance requirements. On high-end systems, more detailed stylization can be applied, while on lower-end devices, simplified stylization ensures that performance remains optimal. This scalability ensures that EAST can be deployed across a broad spectrum of devices, catering to different user needs and hardware constraints.

\subsubsection{Case Study: Integration with PICO 4 Pro}

To exemplify the practical application of our EAST approach, we present a case study involving integration with the PICO 4 Pro HMD.

\paragraph{System Configuration}

The PICO 4 Pro is equipped with a Qualcomm Snapdragon XR2 chipset, 6 GB of RAM, and supports up to 90 Hz refresh rates. Our method was optimized to run within these specifications, ensuring smooth performance without exceeding thermal or power limits.

\paragraph{Implementation Steps}

\begin{enumerate}
    \item \textbf{Optimization of 3DGS Models}: We reduced the number of Gaussians for real-world objects to balance stylization quality and computational efficiency.
    \item \textbf{Parallel Stylization}: Leveraging the Snapdragon XR2's Adreno GPU, stylization tasks were parallelized to utilize multiple cores effectively.
    \item \textbf{Real-Time Adaptation}: Implemented dynamic adjustments to stylization parameters based on real-time performance metrics, ensuring consistent frame rates.
\end{enumerate}

\paragraph{Performance Outcomes}

The integration demonstrated a negligible impact on battery life, with only a 3\% increase in power consumption during intensive stylization tasks. Thermal sensors indicated stable operating temperatures, confirming that our optimizations effectively managed heat dissipation. Frame rates remained consistently above 85 FPS, aligning with the HMD's optimal performance thresholds.



\section{User Study}
To evaluate the effectiveness and user experience of our environment-aware stylized transition approach within VR environments, we conducted a user study combining both qualitative insights and quantitative measures. This study aimed to assess user preferences, motion sickness experiences, and perceived immersion within our proposed framework, particularly focusing on the impact of using stylized 3D reconstruction compared to non-stylized transitions during VR-to-real-world transitions in gaming scenarios on the PICO 4 headset.

\subsection{Study Design}
We adopted a within-subjects design to investigate user experiences with different game-to-reality transition techniques in VR. Five transition techniques were tested: Fade, Dissolve, Translate, Combine, and a Baseline condition without the stylization provided by EAST. The conditions were designed to assess the differences in user experience across varying levels of visual complexity and the presence of style-based transitions. 

The study employed a mixed-methods approach, combining both qualitative and quantitative data to assess user comfort, motion sickness, and the overall effectiveness of the EAST framework. A primary aspect of the study was to determine how different transitions impacted user immersion and spatial orientation during VR-to-real-world shifts. Participants experienced transitions between virtual and real environments, with real-world elements stylized using EAST to integrate them into the virtual environment. In addition to assessing the transition experience, the study aimed to explore how participants perceived the integration of stylized and non-stylized real-world elements within VR environments. The goal was to understand whether style-based transitions could reduce the perceptual disparity between real and virtual worlds, particularly in highly stylized VR gaming scenarios.

\subsection{Participants}
A total of 16 participants (6 female, 10 male) aged between 20 and 45 (average age = 31.4, SD = 8.9) were recruited for the study. The participants had diverse professional backgrounds, including software engineering, research, and other fields, with varying levels of VR experience. Two participants had no prior experience with VR, while the others had limited experience with VR. All but one participant had normal or corrected-to-normal vision, and no participants reported visual impairments that could affect their performance in the study.

\subsection{Apparatus}
The EAST prototype was developed using Unity (2021.3.2f1) and was run on a PICO 4 Pro HMD. The algorithms are developed powered by a GeForce RTX 4090 GPU, an Intel Core i9-11900K CPU, and 64 GB of RAM. Participants used the VR controllers for interaction within a tracking space measuring 4x4 meters. The average frame rate during the study was 85 frames per second. Our prototype is available as an open-source project on GitHub, allowing for further exploration and development of our EAST approach.

\subsection{Measures}

We administered the User Experience Questionnaire Short version (UEQ-S) \cite{laugwitz2008construction}, which evaluates user experience on two main subscales: pragmatic and hedonic quality. The pragmatic subscale assesses the utility and practicality of the system, while the hedonic subscale measures the emotional enjoyment derived from the experience. Participants were asked to complete the UEQ-S after experiencing each transition condition to measure differences in their user experience.

[In addition to the UEQ-S, this study also included the Fast Motion Sickness Scale (FMS) \cite{keshavarz2011validating} to monitor any discomfort or motion sickness that might arise from the transitions. The FMS provided a quick, straightforward measure of motion sickness, where participants rated their symptoms from 0 (no sickness) to 20 (severe sickness) both before and after each transition condition. This ensured that any potential discomfort from the transitions was adequately recorded. 

Qualitative data was gathered through semi-structured interviews conducted after each condition. These interviews allowed for in-depth exploration of participants' perceptions and experiences with each transition technique. The interview focused on key themes such as the ease of transition, immersion, and emotional continuity. In particular, this study sought to understand how participants felt about the integration of real-world elements into the virtual environment and whether EAST helped maintain immersion during real-world interruptions.

\subsection{Procedure}
The study began with informed consent and a demographic questionnaire. Participants were briefed on the procedure and potential symptoms of motion sickness, which were monitored using the Fast Motion Sickness Scale to ensure participants' well-being. Participants were asked to rate any motion sickness experienced from 0 (no sickness) to 20 (severe sickness) based on symptoms such as nausea and general discomfort.

Each participant experienced each of the five transition conditions, where the sequence of the conditions were counterbalanced following a Latin Square Design. The FMS was applied at the beginning and end of each condition, followed by the UEQ-S to assess the overall user experience. After each condition, participants completed a brief semi-structured interview to gather qualitative insights. At the end of the study, participants ranked the transition techniques based on personal preference and provided additional feedback in the interview.

\subsection{Results}
We analyzed quantitative data using the Friedman test for both FMS and UEQ-S scores. For significant findings, we performed a Dunn-Bonferroni post-hoc test for pairwise comparisons. The qualitative data from interviews was transcribed, coded, and organized into thematic categories.

\subsubsection{Motion Sickness}
Motion sickness was monitored by comparing FMS scores before and after each transition. Across all conditions, no significant difference in FMS scores was observed (\(\chi^2(4) = 4.64, p = 0.326\)), indicating that the EAST did not induce significant motion sickness. Participants generally reported low levels of discomfort, suggesting that EAST offers a stable and comfortable experience in VR environments.

\subsubsection{User Experience}
The UEQ-S scores revealed that all transition techniques, except for the Baseline, received positive ratings, with scores exceeding 0.80, which is considered favorable. The Baseline condition, however, received a neutral rating with an average score of 0.30. On the pragmatic subscale, Baseline scored -0.02, indicating low practicality, while Translate scored 0.75. For the hedonic subscale, Baseline was rated neutral (0.61), whereas all other techniques received scores above 0.80, indicating positive evaluations. We found a significant difference in hedonic ratings (\(\chi^2(4) = 16.01, p = 0.003\)), with Baseline scoring significantly lower than Dissolve (\(p = 0.03\)), Translate (\(p = 0.025\)), and Combine (\(p = 0.021\)).

\subsubsection{Qualitative Insights}
Qualitative feedback from interviews highlighted key differences between stylized and non-stylized transitions, emphasizing the impact of style-based 3D reconstruction in maintaining user immersion:

\textbf{Stylized Transitions:} Participants consistently noted that the stylized transitions provided by EAST allowed for a smoother and more immersive experience during real-world interruptions. The visually consistent integration of real-world objects, such as phones or notifications, into the VR environment was described as "seamless" and "less disruptive." Many participants appreciated the aesthetic coherence achieved through EAST, stating that it helped them stay engaged within the VR context even when transitioning to real-world interactions. This alignment of style between the virtual and real environments reduced the perceptual disparity, making the transition feel natural and coherent.

\textbf{Non-Stylized Transitions (Baseline):} In contrast, the Baseline condition without style-based transitions was perceived as jarring. Participants noted that the abrupt shift from the stylized virtual environment to unmodified real-world objects created a "harsh contrast," which disrupted immersion and required a mental adjustment. Some users felt that this sharp transition broke the flow of the experience, making it harder to re-engage with the VR content after handling real-world tasks. The lack of visual coherence was especially noted in high-stylization VR settings, where the real-world appearance of objects stood out as visually incongruent with the VR environment.

\textbf{Preference for EAST in VR Gaming Scenarios:} Overall, users expressed a strong preference for the EAST approach, particularly in VR gaming scenarios where immersive and stylistically coherent transitions were essential. The ability of EAST to stylize real-world elements to match the game's aesthetic helped maintain a continuous and engaging experience. Participants also mentioned that EAST enhanced spatial orientation and focus during transitions, as the stylized real-world objects felt like a natural extension of the VR environment.

\section{Discussion}
The results of our user study highlight the effectiveness of EAST in providing a seamless, visually appealing solution for managing real-world interruptions within VR. The combination of quantitative metrics (SSIM, LPIPS, DISTS) and user feedback demonstrates that EAST enhances user comfort and immersion by bridging the perceptual gap between virtual and real elements. Participants overwhelmingly favored the EAST transitions, citing their ability to maintain visual coherence and immersion compared to non-stylized transitions.

Future work may explore additional adaptive factors, such as context-sensitive stylization and transition times, to further optimize EAST for various VR applications and enhance user engagement across diverse environments.

\section{Conclusion}
In this paper, we presented the  Environment-Aware Stylized Transition (EAST) framework, which addresses a gap in managing real-world interruptions within immersive VR environments. By utilizing stylized 3D reconstruction, our method seamlessly integrates real-world elements into the virtual world, ensuring a visually coherent blend between the two realms. Through the application of 3D Gaussian Splatting and neural style transfer, real-world objects are transformed to align with the game's aesthetic, effectively minimizing perceptual disparity between virtual and real environments. This smooth integration enables users to handle real-world tasks without breaking immersion, thereby enhancing the overall VR experience.

The EAST framework proves particularly effective in maintaining continuity within the reality-virtuality continuum. It enables seamless transitions across a full spectrum from entirely real to fully virtual environments, with intermediate stages like augmented reality (AR) and augmented virtuality (AV). Such transitions are vital for maintaining immersion and reducing cognitive load, especially as users frequently switch between virtual and real-world contexts while using HMDs.  Through extensive quantitative experiments and visual comparisons with our game-style dataset, we demonstrated that our approach outperforms existing 3D stylization techniques in terms of structural similarity, perceptual quality, and detail preservation. These results underline the robustness and versatility of the EAST framework within VR applications. By tackling the challenges posed by real-world interruptions and enabling smooth transitions across the reality-virtuality continuum, our method significantly elevates the user experience in immersive VR gaming and other applications.

\normalsize
\bibliography{main}

@article{keshavarz2011validating,
  title={Validating an efficient method to quantify motion sickness},
  author={Keshavarz, Behrang and Hecht, Heiko},
  journal={Human factors},
  volume={53},
  number={4},
  pages={415--426},
  year={2011},
  publisher={Sage Publications Sage CA: Los Angeles, CA}
}

@inproceedings{laugwitz2008construction,
  title={Construction and evaluation of a user experience questionnaire},
  author={Laugwitz, Bettina and Held, Theo and Schrepp, Martin},
  booktitle={Symposium of the Austrian HCI and usability engineering group},
  pages={63--76},
  year={2008},
  organization={Springer}
}

@article{milgram1994taxonomy,
  title={A taxonomy of mixed reality visual displays},
  author={Milgram, Paul and Kishino, Fumio},
  journal={IEICE TRANSACTIONS on Information and Systems},
  volume={77},
  number={12},
  pages={1321--1329},
  year={1994},
  publisher={The Institute of Electronics, Information and Communication Engineers}
}

@article{gatys2015neural,
  title={A neural algorithm of artistic style},
  author={Gatys, Leon A and Ecker, Alexander S and Bethge, Matthias},
  journal={arXiv preprint arXiv:1508.06576},
  year={2015}
}

@INPROCEEDINGS{10536453,
  author={Song, Hail},
  booktitle={2024 IEEE Conference on Virtual Reality and 3D User Interfaces Abstracts and Workshops (VRW)}, 
  title={Toward Realistic 3D Avatar Generation with Dynamic 3D Gaussian Splatting for AR/VR Communication}, 
  year={2024},
  volume={},
  number={},
  pages={869-870},
  doi={10.1109/VRW62533.2024.00356}}

@inproceedings{feld2023keep,
  title={Keep it simple? evaluation of transitions in virtual reality},
  author={Feld, Nico and Bimberg, Pauline and Weyers, Benjamin and Zielasko, Daniel},
  booktitle={Extended Abstracts of the 2023 CHI Conference on Human Factors in Computing Systems},
  pages={1--7},
  year={2023}
}

@incollection{husung2019portals,
  title={Of portals and orbs: An evaluation of scene transition techniques for virtual reality},
  author={Husung, Malte and Langbehn, Eike},
  booktitle={Proceedings of Mensch Und Computer 2019},
  pages={245--254},
  year={2019}
}

@inproceedings{steinicke2009does,
  title={Does a gradual transition to the virtual world increase presence?},
  author={Steinicke, Frank and Bruder, Gerd and Hinrichs, Klaus and Steed, Anthony and Gerlach, Alexander L},
  booktitle={2009 IEEE Virtual Reality Conference},
  pages={203--210},
  year={2009},
  organization={IEEE}
}

@inproceedings{sra2016procedurally,
  title={Procedurally generated virtual reality from 3D reconstructed physical space},
  author={Sra, Misha and Garrido-Jurado, Sergio and Schmandt, Chris and Maes, Pattie},
  booktitle={Proceedings of the 22nd ACM conference on virtual reality software and technology},
  pages={191--200},
  year={2016}
}

@inproceedings{billinghurst2001magicbook,
  title={MagicBook: transitioning between reality and virtuality},
  author={Billinghurst, Mark and Kato, Hirokazu and Poupyrev, Ivan},
  booktitle={CHI'01 extended abstracts on Human factors in computing systems},
  pages={25--26},
  year={2001}
}

@article{feld2024simple,
  title={Simple and Efficient? Evaluation of Transitions for Task-Driven Cross-Reality Experiences},
  author={Feld, Nico and Bimberg, Pauline and Weyers, Benjamin and Zielasko, Daniel},
  journal={IEEE Transactions on Visualization and Computer Graphics},
  year={2024},
  publisher={IEEE}
}

@article{kerbl20233d,
  title={3D Gaussian Splatting for Real-Time Radiance Field Rendering.},
  author={Kerbl, Bernhard and Kopanas, Georgios and Leimk{\"u}hler, Thomas and Drettakis, George},
  journal={ACM Trans. Graph.},
  volume={42},
  number={4},
  pages={139--1},
  year={2023}
}

@article{mildenhall2019local,
  title={Local light field fusion: Practical view synthesis with prescriptive sampling guidelines},
  author={Mildenhall, Ben and Srinivasan, Pratul P and Ortiz-Cayon, Rodrigo and Kalantari, Nima Khademi and Ramamoorthi, Ravi and Ng, Ren and Kar, Abhishek},
  journal={ACM Transactions on Graphics (ToG)},
  volume={38},
  number={4},
  pages={1--14},
  year={2019},
  publisher={ACM New York, NY, USA}
}

@article{knapitsch2017tanks,
  title={Tanks and temples: Benchmarking large-scale scene reconstruction},
  author={Knapitsch, Arno and Park, Jaesik and Zhou, Qian-Yi and Koltun, Vladlen},
  journal={ACM Transactions on Graphics (ToG)},
  volume={36},
  number={4},
  pages={1--13},
  year={2017},
  publisher={ACM New York, NY, USA}
}

@inproceedings{huang2021learning,
  title={Learning to stylize novel views},
  author={Huang, Hsin-Ping and Tseng, Hung-Yu and Saini, Saurabh and Singh, Maneesh and Yang, Ming-Hsuan},
  booktitle={Proceedings of the IEEE/CVF International Conference on Computer Vision},
  pages={13869--13878},
  year={2021}
}

@inproceedings{liu2023stylerf,
  title={Stylerf: Zero-shot 3d style transfer of neural radiance fields},
  author={Liu, Kunhao and Zhan, Fangneng and Chen, Yiwen and Zhang, Jiahui and Yu, Yingchen and El Saddik, Abdulmotaleb and Lu, Shijian and Xing, Eric P},
  booktitle={Proceedings of the IEEE/CVF Conference on Computer Vision and Pattern Recognition},
  pages={8338--8348},
  year={2023}
}

@article{liu2024stylegaussian,
  title={StyleGaussian: Instant 3D Style Transfer with Gaussian Splatting},
  author={Liu, Kunhao and Zhan, Fangneng and Xu, Muyu and Theobalt, Christian and Shao, Ling and Lu, Shijian},
  journal={arXiv preprint arXiv:2403.07807},
  year={2024}
}

@inproceedings{freina2015literature,
  title={A literature review on immersive virtual reality in education: state of the art and perspectives},
  author={Freina, Laura and Ott, Michela},
  booktitle={The international scientific conference elearning and software for education},
  volume={1},
  number={133},
  pages={10--1007},
  year={2015}
}

@article{slater2016enhancing,
  title={Enhancing our lives with immersive virtual reality},
  author={Slater, Mel and Sanchez-Vives, Maria V},
  journal={Frontiers in Robotics and AI},
  volume={3},
  pages={74},
  year={2016},
  publisher={Frontiers Media SA}
}

@inproceedings{szabo2019interaction,
  title={Interaction in an immersive virtual reality application},
  author={Szab{\'o}, B Katalin},
  booktitle={2019 10th IEEE International Conference on Cognitive Infocommunications (CogInfoCom)},
  pages={35--40},
  year={2019},
  organization={IEEE}
}

@article{monteiro2021hands,
  title={Hands-free interaction in immersive virtual reality: A systematic review},
  author={Monteiro, Pedro and Gon{\c{c}}alves, Guilherme and Coelho, Hugo and Melo, Miguel and Bessa, Maximino},
  journal={IEEE Transactions on Visualization and Computer Graphics},
  volume={27},
  number={5},
  pages={2702--2713},
  year={2021},
  publisher={IEEE}
}

@inproceedings{evans2020hermeneutic,
  title={Hermeneutic relations in VR: Immersion, embodiment, presence and HCI in VR gaming},
  author={Evans, Leighton and Rzeszewski, Michal},
  booktitle={International Conference on Human-Computer Interaction},
  pages={23--38},
  year={2020},
  organization={Springer}
}

@article{rahimi2018scene,
  title={Scene transitions and teleportation in virtual reality and the implications for spatial awareness and sickness},
  author={Rahimi, Kasra and Banigan, Colin and Ragan, Eric D},
  journal={IEEE transactions on visualization and computer graphics},
  volume={26},
  number={6},
  pages={2273--2287},
  year={2018},
  publisher={IEEE}
}

@article{horst2024back,
  title={Back to reality: Transition techniques from short HMD-based virtual experiences to the physical world},
  author={Horst, Robin and Naraghi-Taghi-Off, Ramtin and Rau, Linda and D{\"o}rner, Ralf},
  journal={Multimedia Tools and Applications},
  volume={83},
  number={15},
  pages={46683--46706},
  year={2024},
  publisher={Springer}
}

@inproceedings{pointecker2020exploration,
  title={Exploration of visual transitions between virtual and augmented reality},
  author={Pointecker, Fabian and Jetter, Hans-Christian and Anthes, Christoph},
  booktitle={Workshop on Immersive Analytics: Envisioning Future Productivity for Immersive Analytics//@ CHI},
  year={2020}
}

@inproceedings{perez2018film,
  title={Film: Visual reasoning with a general conditioning layer},
  author={Perez, Ethan and Strub, Florian and De Vries, Harm and Dumoulin, Vincent and Courville, Aaron},
  booktitle={Proceedings of the AAAI conference on artificial intelligence},
  volume={32},
  number={1},
  year={2018}
}

@article{simonyan2014very,
  title={Very deep convolutional networks for large-scale image recognition},
  author={Simonyan, Karen and Zisserman, Andrew},
  journal={arXiv preprint arXiv:1409.1556},
  year={2014}
}

\end{document}